\documentclass[conference]{IEEEtran}
\IEEEoverridecommandlockouts
\usepackage{cite}
\usepackage{amsmath,amssymb,amsfonts}
\usepackage{algorithmic}
\usepackage{graphicx}
\usepackage{textcomp}
\usepackage{xcolor}
\usepackage{stfloats}
\usepackage[switch]{lineno}

\def\BibTeX{{\rm B\kern-.05em{\sc i\kern-.025em b}\kern-.08em
    T\kern-.1667em\lower.7ex\hbox{E}\kern-.125emX}}

\begin{document}
\bstctlcite{IEEEexample:BSTcontrol}

\title{Mitigating the Impact of Retention Loss on Inference Accuracy in 65 nm Single-Poly Floating-Gate Analog In-Memory Computing}

\author{
    \IEEEauthorblockN{
        Mirko Brazzini$^{1}$, 
        Giulio Filippeschi$^{1}$,
        Alessandro Catania$^{1}$, 
        Sebastiano Strangio$^{1}$, 
        Giuseppe Iannaccone$^{1,2}$
    }
    \IEEEauthorblockA{
        $^1$Dipartimento di Ingegneria dell'Informazione, University of Pisa, Pisa, Italy, 
        $^2$Quantavis SRL, Pisa, Italy \\
        Email: mirko.brazzini@phd.unipi.it, giuseppe.iannaccone@unipi.it \thanks{This work has been submitted to the IEEE for possible publication.
Copyright may be transferred without notice, after which this version may no
longer be accessible. This work is partially supported by the EC through the project 2DADDICT (grant n. 101223249) and by the Italian MUR under the Forelab project of the “Dipartimenti di Eccellenza” programme.} 
    }
}

\maketitle

\begin{abstract}
We show with experiments and system-level simulations that it is possible to successfully mitigate the impact of retention loss on inference accuracy degradation by using both circuit-level compensation techniques and batch normalization recalibration at the algorithmic level. Experiments are performed on a single-poly floating-gate (FG) analog non-volatile memory array for analog in-memory computing fabricated in a standard 65 nm CMOS. We use a model of retention-loss statistics calibrated with experiments to evaluate the system-level impact on neural network models such as VGG-10/CIFAR-10 and WideResNet-28-10/CIFAR-100. We show that, after 60 days since programming, combined mitigation techniques enable to recover the baseline inference accuracy within 2-4\%.
\end{abstract}

\begin{figure*}[b]
\centering
\includegraphics[width=\textwidth]{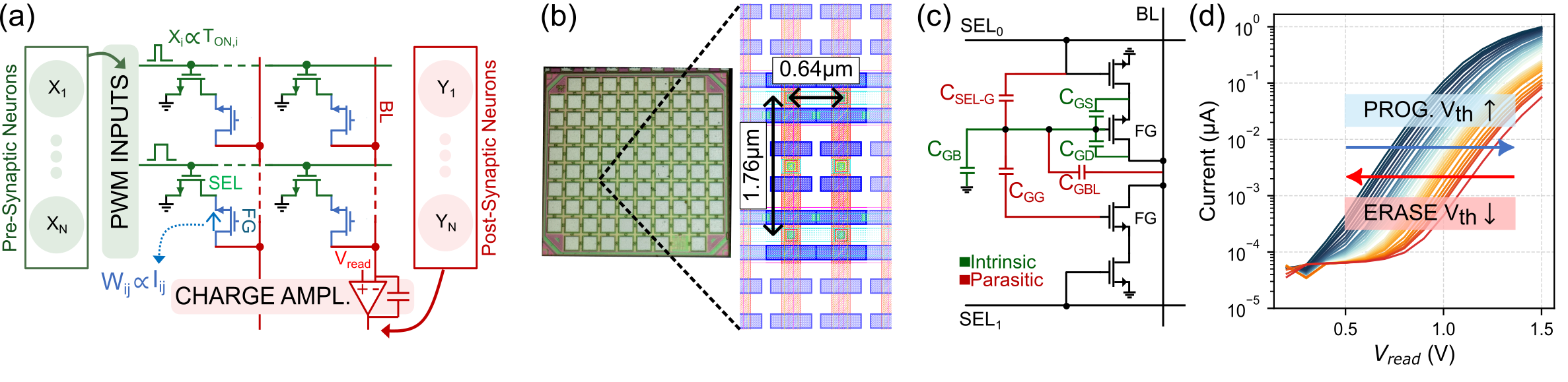}
\vspace{-25pt}
\caption{
(a)~Time-domain analog VMM scheme based on 1T-1FG cells.
(b)~Chip micrograph and layout of a $2{\times}2$ memory cell structure (including dummies).
(c) Equivalent circuit showing the floating-gate potential modulation through the applied $V_{\rm DS}$ and the capacitive coupling network, including $C_{\rm gs}$, $C_{\rm gb}$, $C_{\rm gd}$,  $C_{\rm sel-g}$,  $C_{\rm gg}$ and parasitic capacitances.
(d)~Measured $I$--$V$ characteristics for different programmed conditions.}
\label{fig:panel1}
\vspace{-15pt}

\end{figure*}
\section{Introduction}
Analog vector-matrix multipliers (VMM) for deep learning, where matrix elements (weights) are spatially instantiated in an analog non-volatile memory array enabling Analog In-Memory Computing (AIMC), promise improved energy efficiency by reducing data movement and exploiting subthreshold operation \cite{LeGallo2023,Ambrogio2018,Wan2022,Rizzo2022}. Several memory technologies have been investigated for this application, including phase-change (PCM) \cite{LeGallo2023,Ambrogio2018}, resistive RAM (RRAM) \cite{Wan2022}, and floating-gate (FG) memories \cite{Rizzo2022,Rizzo2026,Agarwal2019,Kim2018,Hasler2016,Mathews2024,Ma2008,Wang2007,Bayat2015,Merrikh2018}. Among these, FG devices offer the advantage of using a mature CMOS process with no additional masks and intrinsic multi-level programmability, at the cost of higher write voltage and reduced endurance \cite{Mehonic2024} \cite{Berggren2021}. For analog non-volatile memories used in deep-learning inference, the impact of retention loss must be assessed in terms of  \textit{inference accuracy degradation}, requiring a device-to-system analysis. Unlike digital non-volatile memories, where the bit error rate is a sufficient and widely accepted metric because cells store a small number of well-separated states protected by error correction, in analog memories each multi-level state directly represents a synaptic weight, so charge loss perturbs the computation itself. For a sufficiently large ensemble of cells, the FG charge loss can be decomposed into a systematic (average) component and a stochastic component that varies randomly and independently from cell to cell, and from one realization to the next, following a common probability distribution. 
By extracting such components from experimental single-poly FG memories fabricated in TSMC 65 nm, we show that circuit-level \cite{Cai2015 ,Luo2018} and algorithmic mitigation techniques can be used to compensate for charge loss and preserve full inference accuracy for more than two months, for different neural network models: an adaptive read voltage, similar to the one used to achieve temperature-resilient neuromorphic chips \cite{Rizzo2026}, compensates for the systematic charge loss component, whereas batch normalization recalibration (BNR) \cite{Joshi2020,Ioffe2015,Bjorck2018} compensates for the zero-average stochastic charge loss component. Using both mechanisms, experimentally calibrated system-level simulations on different neural network architectures show stable inference accuracy after 60 days.
\section{Memory Device and Crossbar Architecture}Crossbar array test structures were fabricated with a single-poly TSMC 65~nm CMOS process using sub-nominal 2.5~V I/O nMOS devices with dimensions of $W/L = 240$~nm$/180$~nm. Considering the VMM scheme in Fig.~\ref{fig:panel1}(a), each 1T-1FG crossbar element consists of a weight-storing FG transistor in series with a selector transistor. The selector transistors eliminate program disturbs for large arrays. The layout and schematic of a $2\times2$ portion of the array are reported in Figs.~\ref{fig:panel1}(b) and (c), respectively, yielding a cell footprint of $0.88~\mu\mathrm{m}\times0.64~\mu\mathrm{m}$. Since we are considering time-domain multiplication \cite {Rizzo2022,Rizzo2026}, each wordline is activated for a time proportional to the corresponding input, turning on the associated selector transistors. Correspondingly, a constant $V_{\mathrm{read}}$ bias is applied to all the bitlines (BLs) so that each FG transistor is biased in saturation with drain-to-source voltage  $V_{\mathrm{DS}} = V_{\mathrm{read}}$, and a gate-to-source voltage $V_{\mathrm{GS}} = k V_{\mathrm{read}} + Q_{\mathrm{FG}} / C_{\Sigma}$, where $C_\Sigma$ is the total floating gate capacitance including intrinsic and parasitic components (Fig.~\ref{fig:panel1}(c)), $Q_{\mathrm{FG}}$ is the total charge injected in the FG \cite {Rizzo2022}, and $k = C_{\rm DS}/C_\Sigma$ is the capacitive coupling factor between drain and floating gate. Each FG transistor acts as a programmable current source with the measured $I$--$V_{\mathrm{read}}$ characteristics shown in Fig.~\ref{fig:panel1}(d), enabling the current to be programmed in a range of over two decades for  $V_{\mathrm{read}} \approx 1~\text{V}$, where the device exhibits a subthreshold exponential behavior.

\begin{figure*}[!b]
\centering
\includegraphics[width=\textwidth]{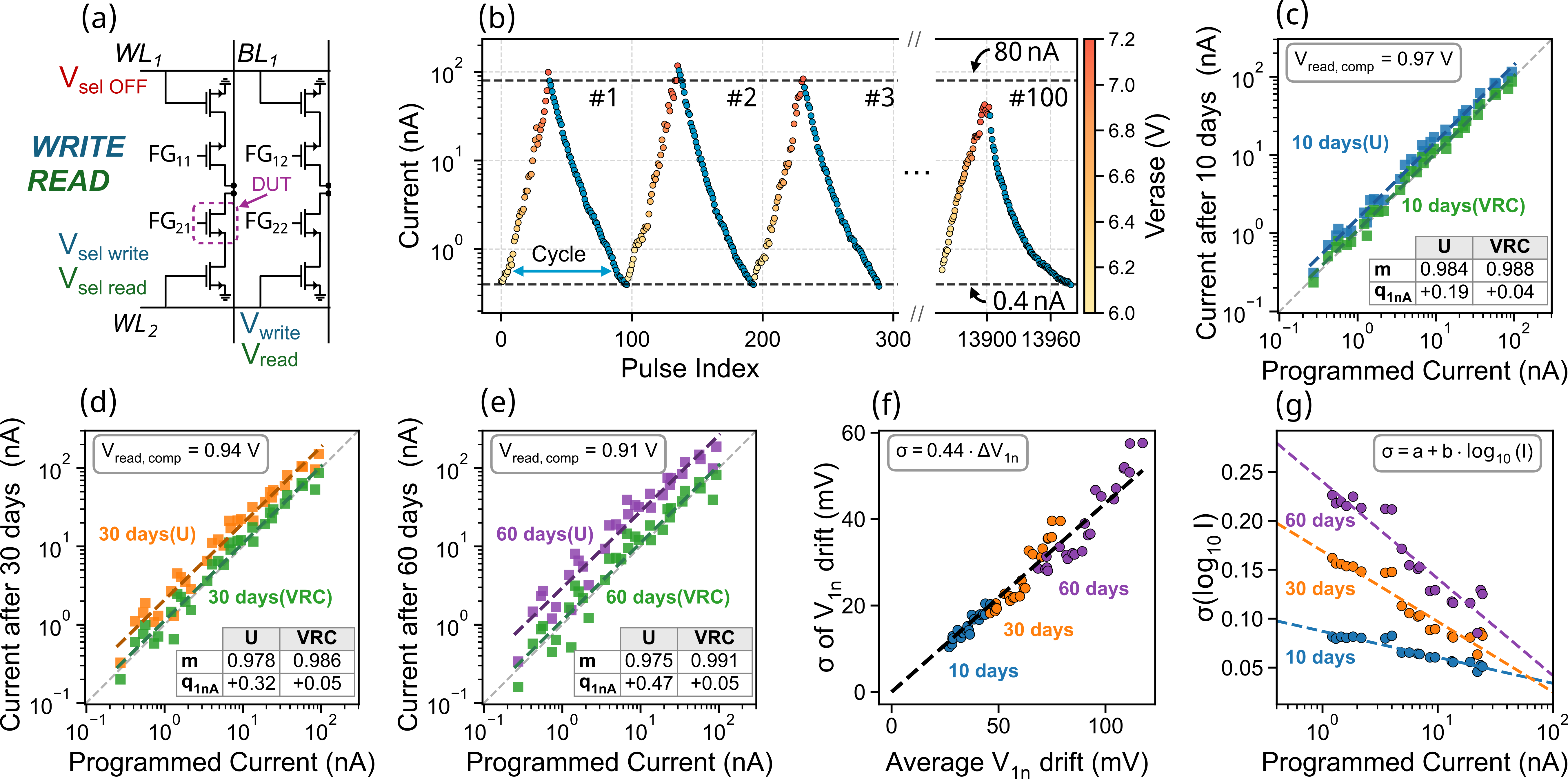}
\vspace{-20pt}
\caption{(a) Bias conditions for program, erase, and read operations. (b) Program/erase cycling characteristics over the first 100 cycles. (c-d-e) Scatter plot of the current measured after 10, 30, and 60 days of room-temperature retention versus the current measured immediately after programming. Dashed lines represent the log--log linear fits obtained for the uncompensated (U) and $V_{\mathrm{read}}$-compensated (VRC) data, from which the retention parameters $m$ (slope) and $q$ (1 nA intercept) are extracted. (f) Standard deviation of the retention-induced drift of
$V_{\rm 1n}$ as a function of the average
$V_{\rm 1n}$ drift. (g) Standard deviation $\sigma (\log_{10}I)$ around the average retention model, expressed in
decades as a function of the programmed current.}

\label{fig:panel2}
\vspace{-10pt}

\end{figure*}

\section{Programming Characteristics and Retention }Fig.~\ref{fig:panel2}(a) illustrates the bias conditions of read and write operations. During read, the BL is held at a constant $V_{\mathrm{read}} = 1~\text{V}$, while the addressed selector is activated with $V_{\mathrm{sel}} = 0.5~\text{V}$ and unselected row selectors are biased at $V_{\mathrm{sel,OFF}} = -1~\text{V}$ to suppress leakage currents, since the selectors are sub-nominal nMOS devices.

Programming and erasing are performed using voltage pulses, which activate channel hot-electron injection (CHEI,
$V_\mathrm{write} = 4.6~\text{V}$, $200~\mu\text{s}$) or impact-ionized hot-hole
injection (IHHI, $V_\mathrm{write} = 6$--$7.2~\text{V}$, $100~\mu\text{s}$), 
respectively~\cite{Rizzo2022,Rizzo2026}.
Fig.~\ref{fig:panel2}(b) shows 100 complete programming cycles,
with a window exceeding two decades. 



Retention is evaluated by monitoring the variation in time of the current of memory cells programmed to distinct levels from $0.4$ to $80~\mathrm{nA}$ during room-temperature storage.
Figure~\ref{fig:panel2}(c,d,e) reports the current measured after 10, 30, and 60 days since programming, respectively, as a function of the current measured immediately after programming. The data are fitted on the log--log scale with a line, with slope $m$ and intercept $q_{1nA}$ (at $1~\mathrm{nA}$). Charge loss from the FG results in a reduction of the transistor threshold voltage and in a corresponding increase of the programmed current. While the average drift can be largely compensated through an appropriate adjustment of $V_{\mathrm{read}}$, the process has significant stochastic and device-to-device variability components.

We can investigate drift and variability by focusing on $V_{1n}$, defined as the applied $V_{\mathrm{read}}$ required to obtain a drain current of $1\,\mathrm{nA}$, which decreases as a consequence of FG charge loss. 
Fig.~\ref{fig:panel2}(f) reports the standard deviation of $V_{1n}$ as a function of the average $V_{1n}$ drift after several days since programming, computed as a moving average over 11 devices: larger standard deviation is observed for
devices exhibiting a larger average drift, with a proportionality that is consistent with the dominant role of device-to-device variability. 

\section{Mitigating the System-Level Impact of Retention Loss}


The average current drift can be modeled using the fitting line on the log-log scale in fig.~\ref{fig:panel2}(e) for 10, 30 and 60 days. Analogously, in fig.~\ref{fig:panel2}(g) we report the standard deviation $\sigma$ of the leakage-induced current drift, expressed in decades of $\log_{10}I$, as a function of the programmed current and of the time since programming. Also in this case, we can model it with a linear function of the logarithm of the programmed current for 10, 30 and 60 days. We use these models into a hardware-aware inference simulation framework \cite{Filippeschi2026}.


\begin{figure}[t]
\centering
\includegraphics[width=\columnwidth]{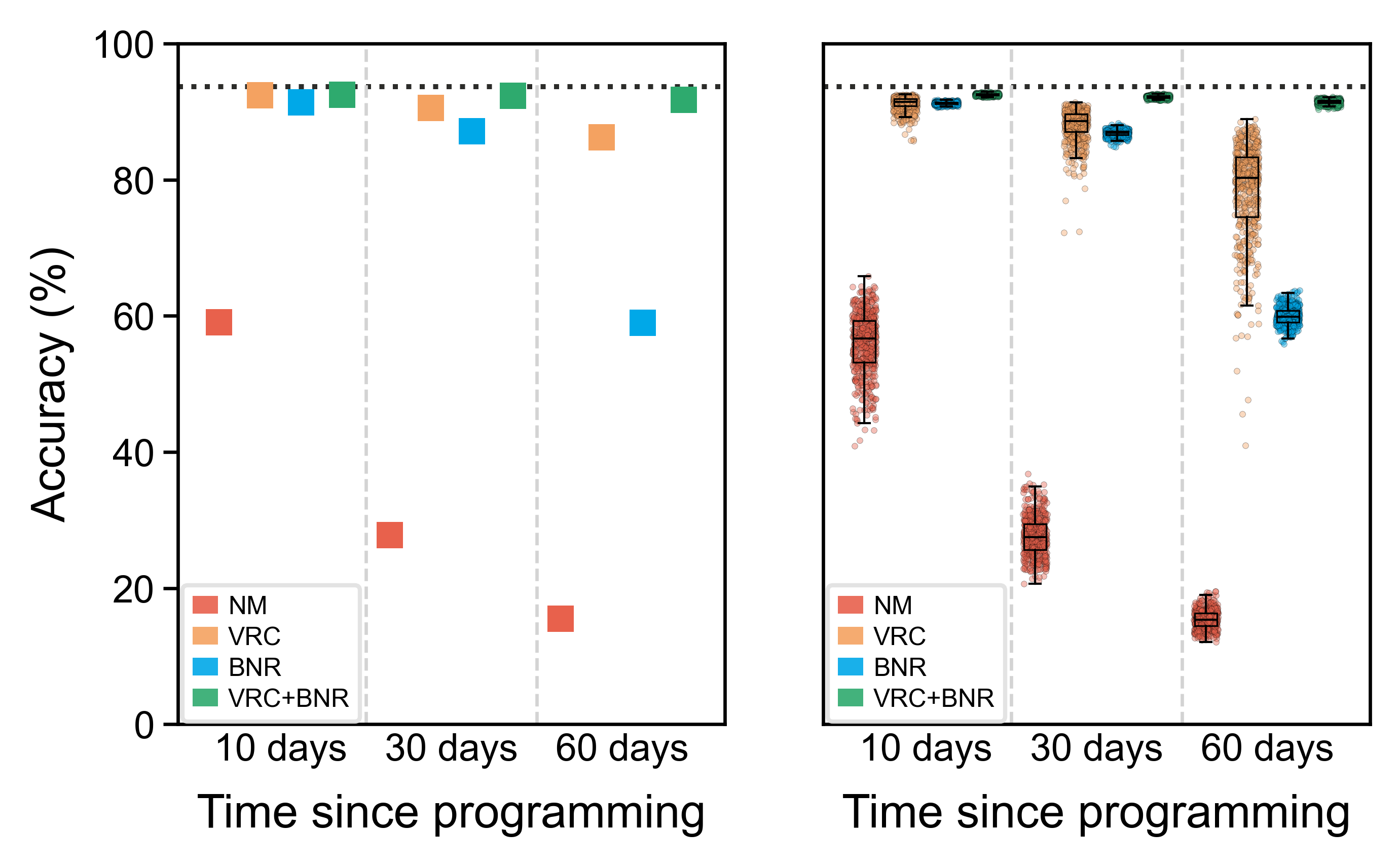}
\vspace{-20pt}
\caption{(a) VGG-10/CIFAR-10 inference accuracy considering only deterministic (average) weight drift. (b) VGG-10/CIFAR-10 inference accuracy for 500 Monte Carlo runs including full stochastic weight drift for four options: no mitigation (NM), $V_{\mathrm{read}}$ compensation (VRC), BNR, and VRC+BNR.}
\label{fig:panel3}
\vspace{-10pt}

\end{figure}

 We report here results for two different types of deep neural networks, among many that we have tested: a VGG-10 network trained on CIFAR-10 and a WideResNet-28-10 trained on CIFAR-100, both including batch-normalization layers. Signed weights are implemented using a differential scheme, where positive and negative weights are mapped onto two independent memory arrays~\cite{Rizzo2026}. The corresponding column currents are subtracted in the analog front-end to reconstruct the signed matrix--vector multiplication. Software weights are mapped onto memory-cell currents according to \begin{equation}
I_{Pi}=\left(\frac{|W_i|}{W_{\max}}\right)\left(I_{\max}-I_{\min}\right)+I_{\min},
\end{equation} where $W_i$ is the $i$-th weight, $I_{Pi}$ is the corresponding programmed current, and $I_{\min} = 0.4\,\mathrm{nA}$ and $I_{\max} = 80\,\mathrm{nA}$ are to the programming window lower and upper bounds. Input activations are quantized to 7 bits. Retention is assessed through the calibrated model \begin{equation}
\log_{10}\!\left[I_i\right]
=
m\cdot\log_{10}\!\left[I_{Pi}\right]
+
q
+
\chi ,
\end{equation} where $\chi = \mathcal{N}(0,\sigma^2(I(t_0))$ is a stochastic variable with a zero-average normal distribution of standard deviation $\sigma$. Of course $m$, $q$ and $\sigma$ depend on the time since programming. 

To evaluate the mitigating effect of $V_{\mathrm{read}}$ compensation and batch normalization recalibration, we consider all four options: $i)$ no mitigation (NM), $ii)$ $V_{\mathrm{read}}$ compensation (VRC), $iii)$ batch-normalization recalibration without $V_{\mathrm{read}}$ compensation (BNR), $iv)$ both VRC and BNR.
 BNR is performed by updating the mean and variance parameters of each batch-normalization layer using a calibration dataset, following the methodology proposed in \cite{Joshi2020}. By re-estimating the activation statistics at the output of each layer, BNR compensates for the distribution shifts caused by retention-induced perturbations in the hardware-mapped weights. Figures~\ref{fig:panel3}(a) and (b) report the inference accuracy of the VGG-10/CIFAR-10 network under different compensation strategies. Figure~\ref{fig:panel3}(a) considers only the average drift component described by $m$ and $q$, whereas Fig.~\ref{fig:panel3}(b) presents data from a Monte Carlo simulation on 500 runs, including the stochastic variability term $\sigma$. Corresponding data are reported in Table~I.
\begin{table}[t]
\caption{Classification accuracy (\%) over time (days).}
\label{tab:retention_transposed}
\centering
\small 
\setlength{\tabcolsep}{5pt} 
\begin{tabular}{l cccc}

 & \multicolumn{2}{c}{\textbf{VGG-10 Ref: 93.7 }} \vspace{2pt}
& \multicolumn{2}{c}{\textbf{WRN-28-10 Ref: 82.1}} \\
\vspace{3pt}

\textbf{Time} & \textbf{BNR} & \textbf{VRC+BNR} & \textbf{BNR} & \textbf{VRC+BNR} \\
\hline
10 & $91.3 \pm 0.2$ & $92.5 \pm 0.1$ & $77.4 \pm 0.3$ & $80.3 \pm 0.2$  \\
30 & $86.9 \pm 0.5$ & $92.2 \pm 0.2$ & $70.8 \pm 0.5$ & $79.6 \pm 0.3$ \\
60 & $59.6 \pm 1.3$ & $91.4 \pm 0.3$ & $45.9 \pm 2.0$ & $78.3 \pm 0.5$ \\
\hline
\end{tabular}
\vspace{-10pt}

\end{table}

\section{Conclusion}

 We have shown the combined effects of $V_{\mathrm{read}}$ compensation and BNR successfully  mitigate the impact of retention loss on inference accuracy for deep neural networks such as VGG-10/CIFAR-10 and WideResNet-28-10/CIFAR-100. After 60 days since programming, the inference is extremely close to the baseline, with very little dispersion. 
 It is important to stress that both mitigation mechanism are required, since $V_{\mathrm{read}}$ compensation acts on the average current drift, whereas BNR acts on the zero-average stochastic component. These techniques provide a practical path toward reliable long-term operation of floating-gate-based analog neuromorphic chips, and will be further investigated for larger neural networks and for longer retention time.

\newpage
\bibliographystyle{IEEEtran}
\bibliography{bibliografia}

\end{document}